\documentclass{ifacconf}

\usepackage{float}
\usepackage{booktabs} 
\usepackage{amsmath,amssymb,amsfonts}
\usepackage{comment,url}
\usepackage{xcolor,bm}
\usepackage{diagbox}
\usepackage{algorithmic}
\usepackage{textcomp}

\usepackage[utf8]{inputenc}
\usepackage{graphicx}
\usepackage{caption}
\usepackage{subcaption}
\usepackage{natbib}
\usepackage{multirow}
\usepackage{url}

\usepackage{natbib}
\usepackage{comment}
\usepackage{colophon}
\begin{document}
\allowdisplaybreaks

\begin{frontmatter}

\title{Balancing a Flying Inverted Pendulum with an Unknown Length Using Model Predictive Control and a Genetic Algorithm Estimator}

\author[First]{Esther Paul} 
\author[First]{Mitchell Torok} 
\author[First]{Mohammad Deghat}

\address[First]{School of Mechanical and Manufacturing Engineering, University of New South Wales (UNSW), 2052, NSW, Australia  (e-mail: e.paul@student.unsw.edu.au, mitchell.torok@student.unsw.edu.au, m.deghat@unsw.edu.au).}

\begin{abstract}
The flying inverted pendulum problem involves balancing an inverted pendulum on an unmanned aerial vehicle (UAV). 
This paper proposes an online Genetic Algorithm (GA) estimator and a Model Predictive Control (MPC) approach to solve the flying inverted pendulum problem in a practical experiment where the pendulum length is unknown. The performance of the MPC approach was demonstrated on a practical system through disturbance and trajectory tracking tests to assess controller robustness and tracking accuracy. The convergence speed and accuracy of the online GA estimator were validated on a practical system using different initial conditions.\footnote{Video of experimental results: \url{https://youtu.be/4zlHNTKoH3Q}}
\end{abstract}


\end{frontmatter}

\section{Introduction}
\begin{figure}[b]
    \centering
    \includegraphics[width=0.9\linewidth]{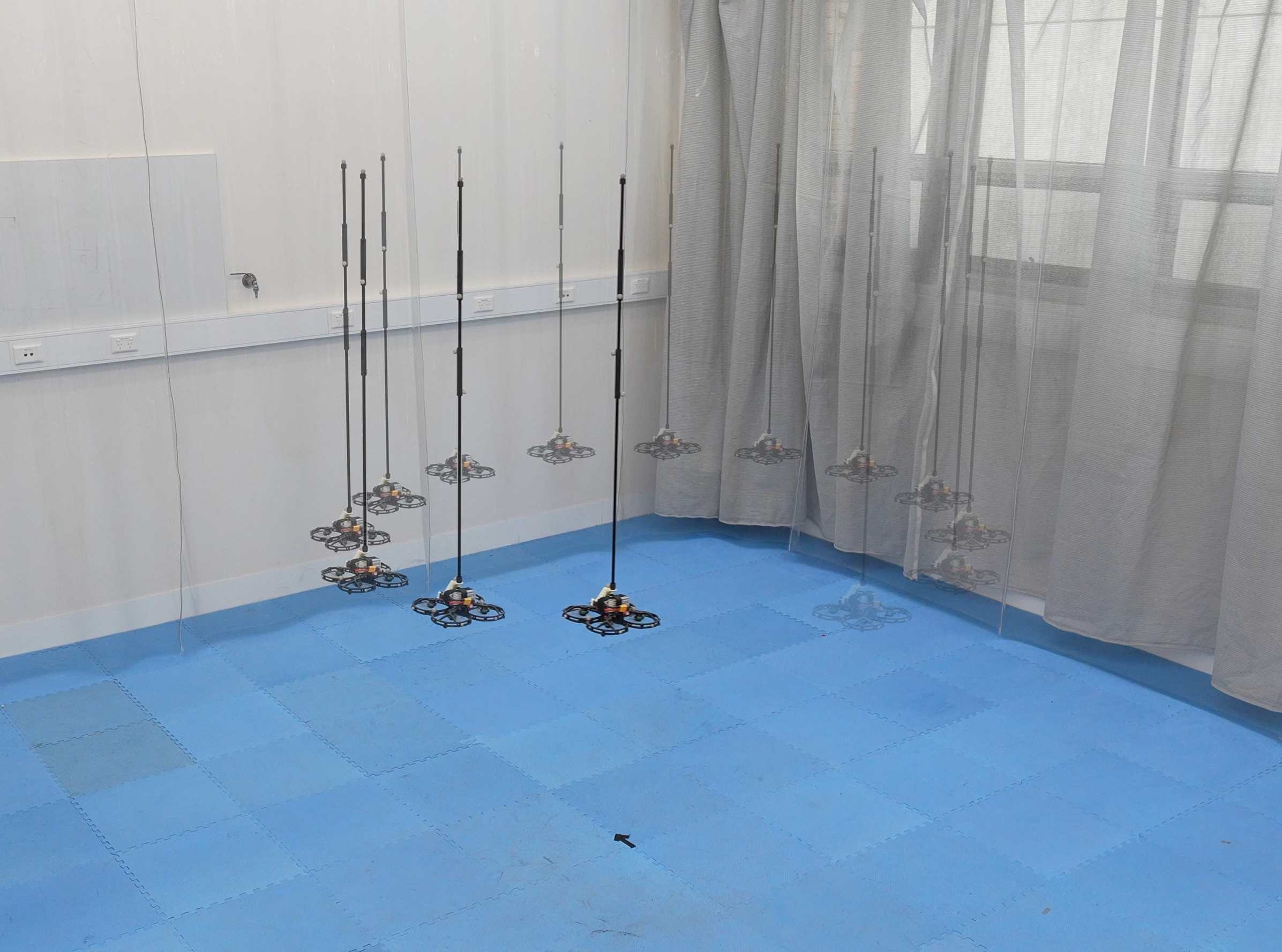} 
    \caption{Flying inverted pendulum with circular trajectory tracking.}
    \label{fig:circle_image}
\end{figure}

The cart inverted pendulum problem has been a benchmark in control theory for decades and is often used to test the effectiveness of a control algorithm.  The Flying Inverted Pendulum (FIP) problem (Fig.~\ref{fig:circle_image}) is an extension of the cart inverted pendulum problem and involves balancing an inverted pendulum on an Unmanned Aerial Vehicle (UAV) instead of a cart. This extension provides insight into the overall stability of UAV dynamics and has potential applications, including stabilisation in vertical takeoff, landing and payload transportation. In addition, the inverted pendulum dynamics is used to model and control rockets in the initial stages of launch \citep{Jannohamed2020FIP}.
The FIP problem poses similar challenges to the cart-inverted pendulum problem involving controlling an underactuated, nonlinear system. Despite these challenges, recent research has shown progress in solving the FIP problem using various control methods. 

\subsection{Model predictive Control}
Model predictive control (MPC) is a control algorithm that chooses optimal control actions based on predictions of the system's behaviour. The algorithm chooses control actions by solving a constrained optimisation problem over a finite number of future time steps known as a control horizon. The selection of optimal control actions over a control horizon is then repeated over a prediction horizon, which is the total time frame the system’s behaviour is predicted for \citep{schwenzer2021mpcReview, mayne2011tube}. The predictive nature of MPC has posed many advantages compared to classical controllers including the ability to improve trajectory tracking,  react to disturbances, and smoothen control actions over time. MPC is also applauded for the ability to handle multi-variable, highly coupled and nonlinear systems while also achieving desired performance by penalizing tracking errors and control actions accordingly. Its ability to impose constraints on system states and control actions also makes it a feasible choice for highly constrained problems \citep{mayne2011tube, yu2024advancedMPC}. 

Due to these advantages, MPC has been investigated as a possible solution to the FIP problem; however,  there is minimal research available regarding this approach. Of the few studies available, \cite{tamba2020regulation} formulated an MPC controller that stabilised the pendulum while moving the quadcopter to a desired position and only validated this controller in simulations. \cite{oloo2023effect} investigated the use of linear adaptive MPC to stabilise a pendulum on a quadcopter with hindered rotor performance, where simulation results showed that the adaptive MPC controller outperformed an LQR controller when stabilising the system. The work presented by \cite{yang2024adaptiveMPC} and \cite{kang2024rrtstar} used nonlinear MPC approaches, which showed good tracking performance and fast error convergence in simulated experiments. This work highlighted the effectiveness of MPC in a simulated setting, however, the feasibility of MPC on real-world hardware is an important consideration since large model mismatches severely affect the control performance \citep{schwenzer2021mpcReview}. To handle model mismatches, an adaptive MPC approach can be used to estimate system parameters. \cite{sferrazza2020learningmpc} proposed a learning-based MPC  framework that focused on alleviating the computational burden of the optimisation problem and improving the accuracy of trajectory tracking by parameterising the MPC prediction model and then learning unmodeled dynamics and repeated disturbances through offline iterative learning trials.

The results of this study indicated that the MPC framework was a feasible solution to the FIP problem in a practical setting; however, there are limited studies available on using a standard MPC controller on a practical FIP system. Since the performance of the MPC controller is dependent on the accuracy of the prediction model, it is worth considering the effects of unmatched uncertainty and how this may be mitigated through online estimation in a practical scenario.

\subsection{Parameter Estimation}
Parametric uncertainties in the highly coupled FIP system dynamics greatly affect the performance of a controller. Model-based controllers have been developed to mitigate these uncertainties. \cite{barawkar2023unknownlength} focused on formulating a controller for an inverted pendulum-quadcopter system where the length of the inverted pendulum was unknown. Concurrent Learning (CL) Model Reference Adaptive Control (MRAC) algorithm was formulated to estimate the unmatched certainty, and the performance of the control algorithm was validated in a simulation.  Similarly,   \cite{yang2024adaptiveMPC} integrated an L1 adaptive controller into their MPC approach to estimate model uncertainties mainly concerning the pendulum length and mass. The simulated results showed that the proposed controller outperformed other tested controllers including a non adaptive MPC controller, an LQR controller and a linear feedback controller. These works showed promising simulation results as the proposed estimators accurately estimated unknown parameters, however, the experimental demonstrations have not been shown, where measurement noise and unmodeled dynamics may impact the estimator's performance, and the controller may face challenges involving computational efficiency.  

\subsection{Genetic Algorithm}
The Genetic Algorithm (GA) is an optimisation approach often used for search-based applications where optimal parameters must be found given a search space and a fitness function.  The GA converges to optimal parameters by emulating the fundamental principles of evolution, including selection, crossover, mutation and natural selection through survival of the fittest \citep{lambora2019gaReview}. Genetic algorithms pose many advantages, including the ability to handle noisy data,  requiring little domain knowledge and avoiding convergence to local minima. However, the main advantage of GAs lies in the ability to adapt to multiple use cases while maintaining strong performance when searching for optimal parameters \citep{Iglesias2023}. Genetic Algorithms have been used for applications ranging from controller parameter tuning to system identification. \cite{chen2019improvedGA} used an Improved Genetic Algorithm (IGA) to select optimal gains for a Linear Quadratic Regulator (LQR) controller to increase the speed of the FIP system response and to reduce the trajectory error. It was found that the quadrotor was able to balance an inverted pendulum with an initial angle offset while following a circular trajectory with high accuracy, indicating the ability of the algorithm to handle the highly coupled dynamics of the FIP system.  Additionally,  \cite{yang2014systemid} used an offline  Genetic Algorithm to identify system parameters such as inertia, lift coefficient, and torque coefficient of a quadcopter system. The results showed that the Genetic Algorithm was able to identify system parameters with high accuracy using flight data obtained through practical experiments. Furthermore,  \cite{srivastava2019aeroGA} used a Genetic Algorithm to estimate longitudinal aerodynamic derivatives with high accuracy using both simulated and experimental data.  The results of the aforementioned studies demonstrated the strong ability of the GA to select optimal control gains in high-dimensional problems with noisy state information.

\subsection{Contributions}
Few studies have tested MPC on a practical FIP system. Additionally, parametric unmatched uncertainty has a large effect on the formulation of control actions, but limited research has been conducted on estimating the unmatched uncertainty, such as the length of the pendulum in the FIP system. To the author's knowledge, no research has validated a pendulum length estimator on a practical flying inverted pendulum system, prone to measurement noise, disturbances and unmodeled dynamics. 
The main contributions of this paper are the design and implementation of an MPC controller to solve the FIP problem on a practical FIP system, and the development and implementation of a Genetic Algorithm–based estimator for real-time pendulum-length estimation on the same system.


\section{Dynamics and Control Design}
This section presents the dynamic model of the flying inverted pendulum (FIP), which is subsequently used for MPC controller design and pendulum length estimation. The model is formulated under two simplifying assumptions. Firstly, the pendulum mass is small relative to the quadcopter mass, such that the pendulum motion does not appreciably perturb the quadcopter dynamics. Secondly, the pendulum base is resting at the geometric centre of the quadcopter, with zero offset in the body-frame \(x\)–\(y\) plane.

\subsection{Quadcopter Dynamic Model}
The quadcopter pose is described by its position and orientation in the inertial frame. The rotation of the body frame with respect to the inertial frame is defined by a set of Euler angles \((\phi,\theta, \psi\)), where $\phi$ is the roll angle, $\theta$ is the pitch angle, and $\psi$ is the yaw angle. The rotation of the body frame with respect to the inertial frame is given by the rotation matrix
\begin{equation}
    R(\phi, \theta, \psi) = R_x(\phi)R_y(\theta)R_z(\psi),
\end{equation}
where $R_x(\phi)$, $R_y(\theta)$, and $R_z(\psi)$ are the standard rotation matrices corresponding to rotations about the body $x-$, $y-$, and $z-$axes, respectively. The translational acceleration  of the quadcopter in the inertial frame is given by the dynamic model
\begin{equation}
\begin{bmatrix}
\ddot{x} \\ \ddot{y} \\ \ddot{z}
\end{bmatrix}
=
R(\phi,\theta,\psi)
\begin{bmatrix}
0 \\ 0 \\ F/m
\end{bmatrix}
-
\begin{bmatrix}
0 \\ 0 \\ g
\end{bmatrix}.
\end{equation}
where \(F\) is the collective force produced by the rotors in the body frame, \(m\) is the mass of the quadcopter and \(g\) is the acceleration due to gravity.

The relationship between the body–frame angular velocity 
\(\boldsymbol{\omega}_b = [\,\omega_x\;\omega_y\;\omega_z\,]^{\top}\)
and the Euler-angle rates is given by the standard kinematic mapping
\begin{equation}
\begin{bmatrix}
\dot{\phi} \\[2pt]
\dot{\theta} \\[2pt]
\dot{\psi}
\end{bmatrix}
=
\begin{bmatrix}
1 & \sin\phi\tan\theta & \cos\phi\tan\theta \\
0 & \cos\phi           & -\sin\phi          \\
0 & \sin\phi/\cos\theta & \cos\phi/\cos\theta
\end{bmatrix}
\begin{bmatrix}
\omega_x \\[2pt]
\omega_y \\[2pt]
\omega_z
\end{bmatrix}.
\end{equation}

\subsubsection{Pendulum System Dynamics:}
The pendulum dynamics were derived using the Lagrangian method. The position of the pendulum’s center of mass (COM) is denoted as 
\begin{equation}
\bm{p_p} = \begin{bmatrix}
     a & b & \zeta
\end{bmatrix}^{\top} \in \mathbb{R}^{3},
\end{equation}
where \(a\),  \(b\), and  \(\zeta\) represent the positions of the pendulum's COM with respect to the base of the quadcopter along the \(x\), \(y\) and \(z\) axes of the inertial frame. This Cartesian parameterisation is adopted for convenience, as it directly enforces the spherical constraint of the rigid pendulum. The vertical coordinate \(\zeta\) is defined as
\begin{equation}
\zeta = \sqrt{\left(\frac{l}{2}\right)^2 - a^2 - b^2}.
\end{equation}
where $l$ denotes the length of the pendulum.
\noindent Given that the quadcopter position in the inertial frame is  \(\left[
x,~ y,~ z
\right]^{\top}\)\!, the global position of the pendulum is \(
\begin{bmatrix}
a + x, & 
y + b, &
\zeta + z
\end{bmatrix}^{\top}\!.
\)
The Lagrangian is given by \( L = K - P \), where $K$ and $P$ are the kinetic and potential energy at the the global position of the pendulum's COM, respectively. Hence, the nonlinear pendulum dynamics can be derived by Lagrange’s equations
\begin{equation}
\frac{d}{dt}\left(\frac{\partial L}{\partial \big(\dot{a},\dot{b}\big)}\right)
- \frac{\partial L}{\partial(a,b)} = 0.
\end{equation}

\subsection{Model Predictive Control}
Model predictive control (MPC) is an optimisation-based approach that computes a control input \(u_k\) at each discrete timestep \(k\) by minimising a cost function over a prediction horizon of length \(N\) over a specified time frame $t_f$. In this work, the MPC regulates the flying inverted pendulum by driving \(a\) and \(b\) towards zero while ensuring the quadcopter tracks the predefined reference trajectory. Only the roll and pitch angular rates are controlled by the MPC. Altitude and yaw are regulated by independent PID controllers and, therefore, excluded from the optimisation, as these states evolve slowly relative to the lateral pendulum motion and do not appreciably influence the FIP dynamics.

To reduce the computational burden associated with the full nonlinear model, the coupled quadcopter--pendulum dynamics are linearised around the hover equilibrium. Only the small-angle behaviour is required for control. Following \cite{hehn2011flying}, and linearising about \(a=b=\phi=\theta=0\) while retaining first-order terms, the relevant \(x\)--\(y\) quadcopter and pendulum dynamics reduce to
\begin{align}
\ddot{x} &= g\,\theta, \\
\ddot{y} &= -g\,\phi, \\
\ddot{a} &= \frac{2ag}{l} - g\,\theta, \\
\ddot{b} &= \frac{2bg}{l} + g\,\phi, \\
\dot{\phi} &= \omega_x, \\
\dot{\theta} &= \omega_y.
\end{align}
The system state and control input are defined as 
\begin{equation}
\mathbf{x} =
\begin{bmatrix}
x & \dot{x} & y & \dot{y} & a & \dot{a} & b & \dot{b} & \phi & \theta
\end{bmatrix}^{\top}
\in \mathbb{R}^{10},
\end{equation}
\begin{equation}
\mathbf{u} =
\begin{bmatrix}
\omega_x & \omega_y
\end{bmatrix}^{\top} \in \mathbb{R}^{2}.
\end{equation}

Since the FIP dynamics are continuous, a fourth-order Runge--Kutta (RK4) method is
used to obtain the discrete-time prediction model required for MPC. The continuous
system
\begin{equation}
\dot{\mathbf{x}} = f(\mathbf{x}, \mathbf{u}, l ),
\label{eq:prediction_model}
\end{equation}
where $l$ denotes the pendulum length, is discretised with sampling time
$\Delta t$ to yield the prediction model
\begin{equation}
\mathbf{x}_{k+1}
= f_{\mathrm{RK4}}\!\left(
    \mathbf{x}_{k},\,
    \mathbf{u}_{k},\,
    l,\,
    \Delta t
\right).
\end{equation}

The reference trajectory $\mathbf{x}^{\mathrm{ref}}_k \in \mathbb{R}^{10}$ is defined
over the MPC prediction horizon, and the tracking error at each timestep is
\begin{equation}
\mathbf{e}_k = \mathbf{x}_k - \mathbf{x}^{\mathrm{ref}}_k .
\end{equation}
The MPC optimisation problem is then formulated as
\begin{equation}
\label{eqn:mpc_main}
\begin{aligned}
\min_{\{ \mathbf{u}_k\}_{k=0}^{N-1}} \quad &
\mathbf{e}_N^\top P \mathbf{e}_N
+ \sum_{k=0}^{N-1} \left(
    \mathbf{e}_k^\top Q \mathbf{e}_k
    + \mathbf{u}_k^\top R \mathbf{u}_k
\right) \\[4pt]
\text{subject to} \quad &
\mathbf{x}_0 = \mathbf{x}, \\[3pt]
& \mathbf{x}_{k+1}
    = f_{\mathrm{RK4}}\!\left(
        \mathbf{x}_k,\,
        \mathbf{u}_k,\,
        l,\,
        \Delta t
      \right), \\[3pt]
& \mathbf{u}_{\min} \leq \mathbf{u}_k \leq \mathbf{u}_{\max}, \\[3pt]
& \mathbf{x}_{\min} \leq \mathbf{x}_k \leq \mathbf{x}_{\max}.
\end{aligned}
\end{equation}
Here, the predefined cost matrices \(P \in \mathbb{R}^{10\times 10}\) and  \(Q \in \mathbb{R}^{10\times 10}\) penalise the state tracking error,  while \(R \in \mathbb{R}^{2\times 2}\) penalises control effort. The bounds
$\mathbf{u}_{\min}, \mathbf{u}_{\max}\in \mathbb{R}^{2}$ and $\mathbf{x}_{\min}, \mathbf{x}_{\max}\in \mathbb{R}^{10}$  are applied element-wise. 
A terminal cost constraint is omitted from the MPC problem formulation since it limits the feasible region of solutions which may cause the MPC solver to fail when the system experiences external disturbances or fluctuations in the estimation of the pendulum length.

\subsection{Genetic Algorithm}
The objective of the Genetic Algorithm (GA) is to estimate the unknown pendulum length $l$ online. The MPC initially operates using a conservative nominal value of $l$. A sliding window of length $t_w$ is used, within which the state and input data $\{(\mathbf{x}_k,\mathbf{u}_k)\}_{k=1}^{t_w}$ are gathered as the system evolves. After this window is filled, the GA is invoked to identify the value of $l$ that best matches the observed motion. This approach requires the quadcopter to follow a time-varying trajectory that excites the system for parameter estimation.

\subsubsection{Population and Fitness Evaluation:}
A population of $N_c$ candidate pendulum lengths is sampled uniformly from the feasible interval $[l_{\min},\, l_{\max}]$. For each candidate value $l_c$, the system dynamics are propagated over the stored data window using a Kalman Filter (KF). This KF provides a model-consistent state estimate under the hypothesis that the true pendulum length equals the candidate $l_c$.

Let $\mathbf{x}_k$ denote the measured state and $\mathbf{u}_k$ the applied input at timestep $k$. For a given candidate $l_c$, the KF maintains a predicted state $\hat{\mathbf{x}}_k$ with covariance $P_k$. The prediction step is
\begin{align}
\hat{\mathbf{x}}_{k+1} &= \hat{\mathbf{x}}_{k} +
    f(\hat{\mathbf{x}}_{k},\mathbf{u}_{k},l_c)\cdot dt, \\
P_{k+1} &= J_x P_k J_x^\top + Q_e .
\end{align}
where $J_x = \partial f / \partial \mathbf{x}$ and $Q_e$ is the process noise covariance and $f$ represents the system dynamics from \eqref{eq:prediction_model}. The measurement residual $\mathbf{y}_k$, measurement prediction covariance $S_k$ and filter gain $K_k$ are calculated as 
\begin{align}
\mathbf{y}_k &= \mathbf{x}_k - H\hat{\mathbf{x}}_k, \\
S_k &= H P_k H^\top + R_e \\
K_k &= P_k H^\top S_k^{-1} 
\end{align}
where $H$ is the measurement model and  $R_e$ is the measurement noise covariance.
The updated state estimate and covariance are given by
\begin{align}
\hat{\mathbf{x}}_k^{+} &= \hat{\mathbf{x}}_k + K_k \mathbf{y}_k, \\
P_k^{+} &= (I - K_k H) P_k .
\end{align}
The fitness of a candidate parameter $l_c$ is defined using the KF innovation, which quantifies the discrepancy between the measured state $\mathbf{x}_k$ and the posterior KF estimate $\hat{\mathbf{x}}_k^{+}$. The accumulated squared innovation over the window of length $t_w$ yields a likelihood-like score
\begin{equation}
\mathrm{fitness}(l)
= \sum_{k=1}^{t_w}
\left\| \mathbf{x}_k - \hat{\mathbf{x}}_{k}^{+} \right\|^{2}.
\end{equation}
Minimising this expression selects the candidate value of $l$ whose predicted dynamics best match the observed system behaviour over the window.

\subsubsection{Selection, Crossover, and Mutation:}
A predefined fraction $r_p$ of the best-performing candidates is preserved for the next generation. The remaining parents are selected using roulette-wheel selection. Each candidate $l_c$ is encoded into an $n$-bit chromosome (with $n$ chosen to cover the interval $[l_{\min},l_{\max}]$ at a resolution of $10^{-4}$). Offspring are generated using single-point crossover and bit-flip mutation with the predefined rate $r_m$. After each iteration the best estimate of $l$ is recorded. A median filter is applied across successive GA solutions to enhance robustness, and the resulting length estimate is dynamically utilised in the model employed by the MPC controller.

\section{Experimental setup}
\begin{figure}[htbp]
    \centering
    \includegraphics[width=0.8\linewidth, trim={0 0 0 0}, clip]{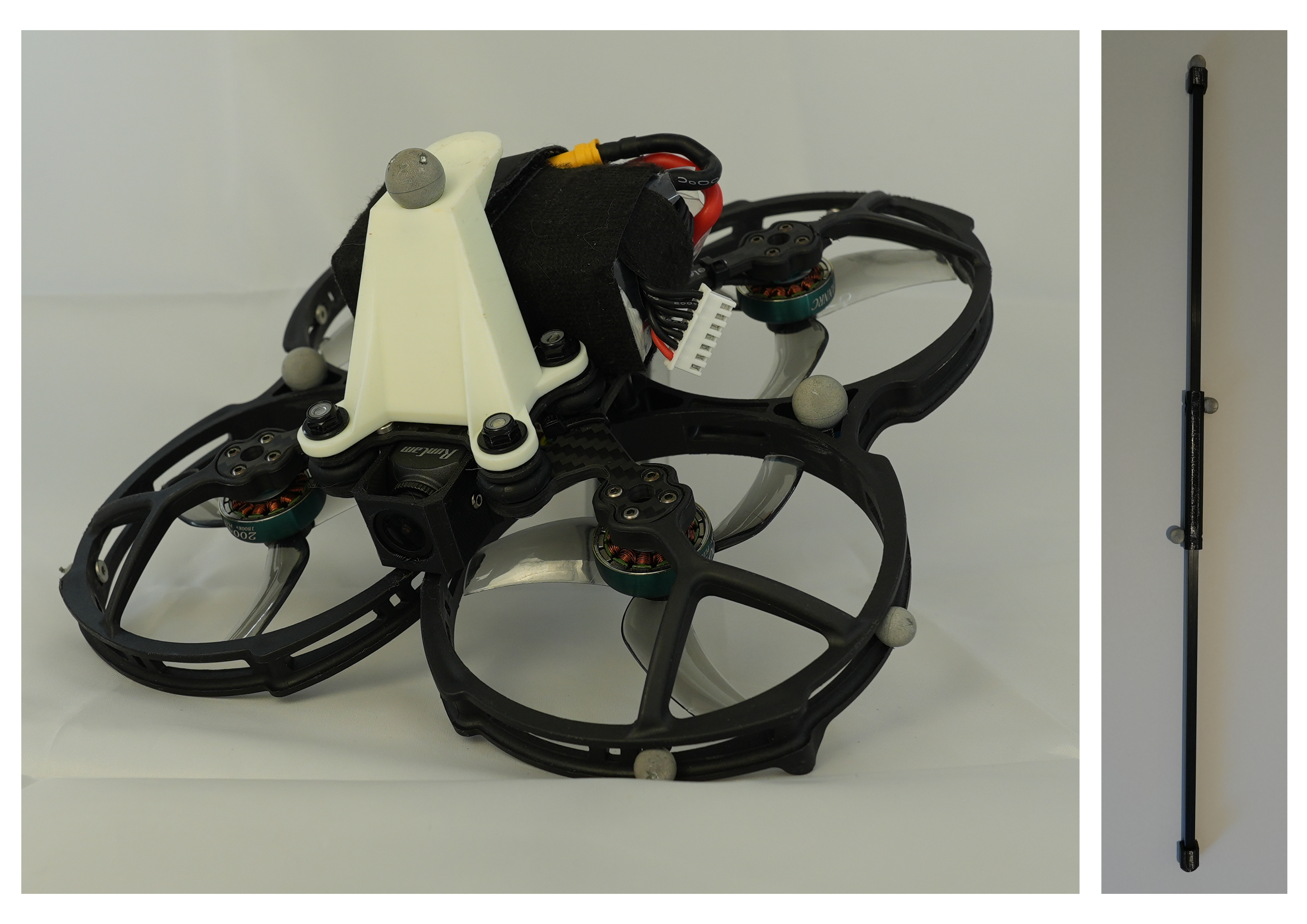}
    \caption{Quadcopter and pendulum used in the experiments. The pendulum sits in a shallow recess above the battery within the 3D-printed structure. }
    \label{fig:quad_image}
\end{figure}

The quadcopter (Fig.~\ref{fig:quad_image}) used a 3.5'' Cinewhoop frame with \(2004\, (1800\,\mathrm{kV})\) motors, a \(45\,\mathrm{A}\) F7 AIO flight controller, and a \(6\mathrm{S}\), \(1300\,\mathrm{mAh}\) LiPo battery. It ran standard Betaflight firmware with a linear angular-rate response and received throttle and desired rate commands from a remote computer via ExpressLRS. The inverted pendulum was a carbon-fibre rod with 3D-printed end caps and tracking-marker mounts, seated in a shallow recess \(8\,\mathrm{cm}\) above the quadcopter frame. The pendulum was placed onto the vehicle once stable hover was achieved, after which the FIP controller was activated. System parameters are listed in Table \ref{tab:params}. Ten OptiTrack cameras operating at \(240\,\mathrm{Hz}\) provided pose estimates, with velocities obtained by differentiating and low-pass filtering the measurements.

\begin{table}[h]
\centering
\caption{Parameters of the quadcopter and pendulum.}
\label{tab:params}
\resizebox{0.9\linewidth}{!}{
\begin{tabular}{llll}
\toprule
\textbf{Symbol} & \textbf{Definition} & \textbf{Value} & \textbf{Unit} \\
\midrule
$m$                 & Mass of quadcopter                    & 0.569           & $kg$ \\
$m_p$               & Mass of pendulum                      & 0.032           & $kg$ \\
$l$                 & Length of pendulum                    & 1.0           & $m$ \\
$\omega_{body}$       & Max quadcopter angular velocity      & 100          & $deg/s$ \\
\bottomrule
\end{tabular}
}
\end{table}

\begin{table}[h]
\centering
\caption{Software implementation parameters.}
\label{tab:software_params}
\resizebox{0.9\linewidth}{!}{
\begin{tabular}{llll}
\toprule
\textbf{Parameter} & \textbf{Definition} & \textbf{Value} & \textbf{unit} \\
\midrule
$N$           & MPC horizon nodes           & 60   &         \\
$t_f$         & MPC time frame        & 6 & $s$            \\
$t_w$         & GA window size       & 55 &    \\
$l_min$       & Minimum pendulum length      & 0.2 & $m$ \\
$l_max$       & Maximum pendulum length      & 2.0 & $m$\\
$r_p$       & GA best performing rate      & 0.167 &\\
$r_m$       & GA bit flip mutation rate    & 0.1 &\\
\bottomrule
\end{tabular}
}
\end{table}

The proposed controller was implemented in Python using the ROS2 framework. The continuous system dynamics were implemented in CasADi, with acados used to solve the optimal control problem at a fixed frequency of $120$Hz. Partial condensing HPIPM was used to reduce the size of the quadratic problem, allowing the problem to be solved in real time. The Genetic Algorithm operated in a separate node so as not to stall the MPC controller. The GA node operated at $0.2$Hz, and sent the pendulum length estimation directly to the MPC controller after each iteration. The system parameters used can be seen in Table \ref{tab:software_params}. The complete system ran on a remote laptop, with a Intel® Core™ i7-240H x 16 CPU and $32$ GB of Memory.

\section{Experimental Results}
The performance of the proposed approach was demonstrated by evaluating the system's response to external disturbance, circular-trajectory tracking, and the convergence of the Genetic Algorithm estimator across varying pendulum lengths.

\subsection{External Disturbance}
The MPC controller performance was tested in the case where the system was subjected to a disturbance while the quadcopter hovered at a position of $(x, y, z) = (0, 0, 1.3m)$ and tried to maintain a vertical pendulum with $a=b = 0$. A disturbance from an external force was applied using a plastic pole at \(5.3\), \(12\), and \(20.2\) seconds.
 
\begin{figure}[htbp] 
    \centering
    \includegraphics[width=0.8\linewidth]{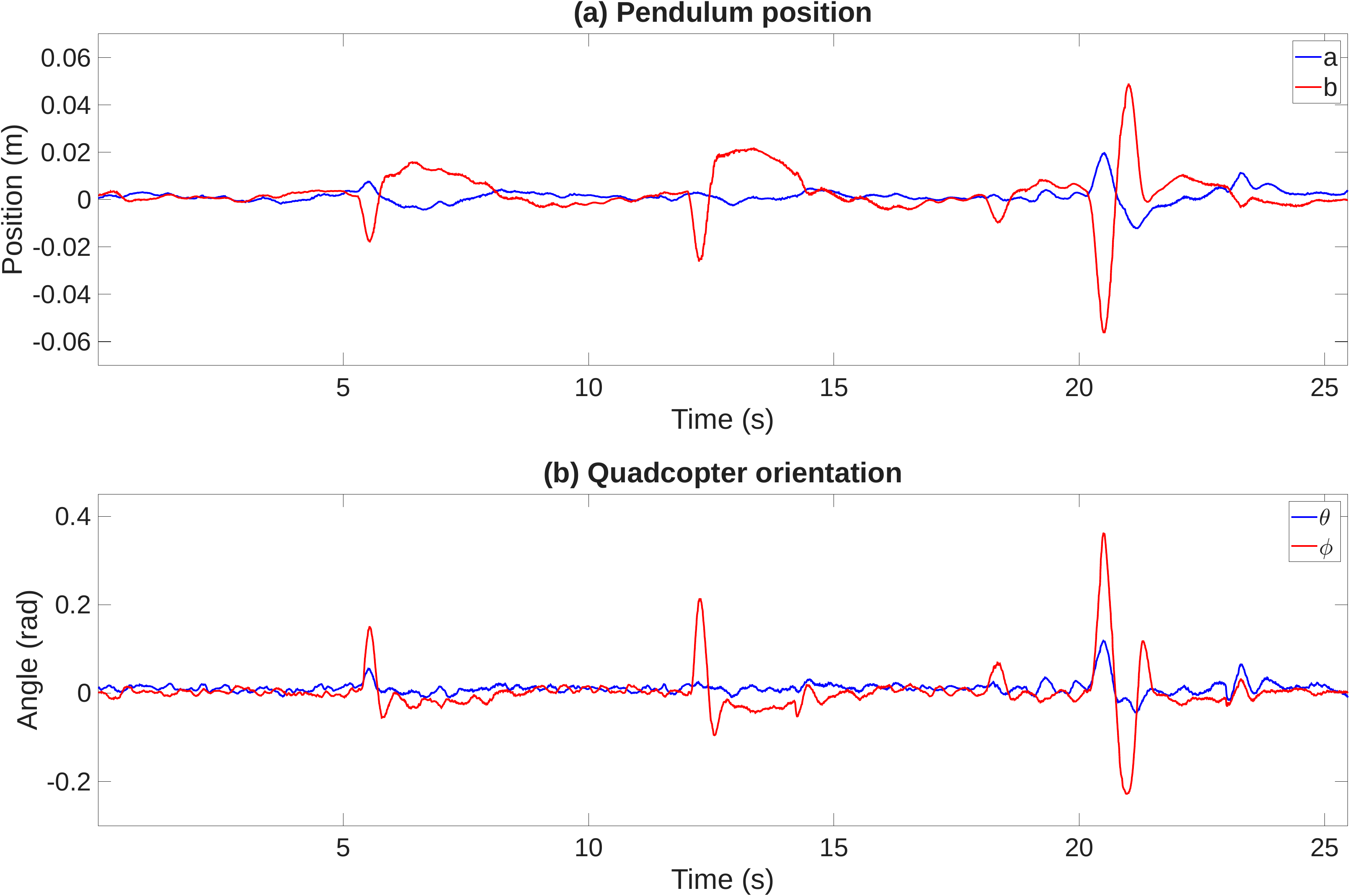}
    \caption{MPC controller performance when subjected to external force disturbances.}
    \label{fig:disturbance}
\end{figure}

The system demonstrated stability in the presence of disturbance. Fig.~\ref{fig:disturbance} showed a fast response to disturbance for the pendulum states with \(a\) and \(b\) showing a settling time of approximately $2$ seconds and a maximum deviation of approximately $5.8$ cm, which translates to an angle of $6.67$ degrees from the vertical position. The quadcopter's ability to continue to balance the inverted pendulum despite disturbances demonstrates the MPC controller's robustness against disturbance. 

\subsection{Circular-trajectory}

 \noindent The MPC controller was tested for its ability to maintain the stability of the FIP system as it moved along a circular trajectory with a radius of $0.8$ meters as seen in Fig.~\ref{fig:circle_image}.
 
\begin{figure}[htbp] 
    \centering
    \includegraphics[width=0.9\linewidth]{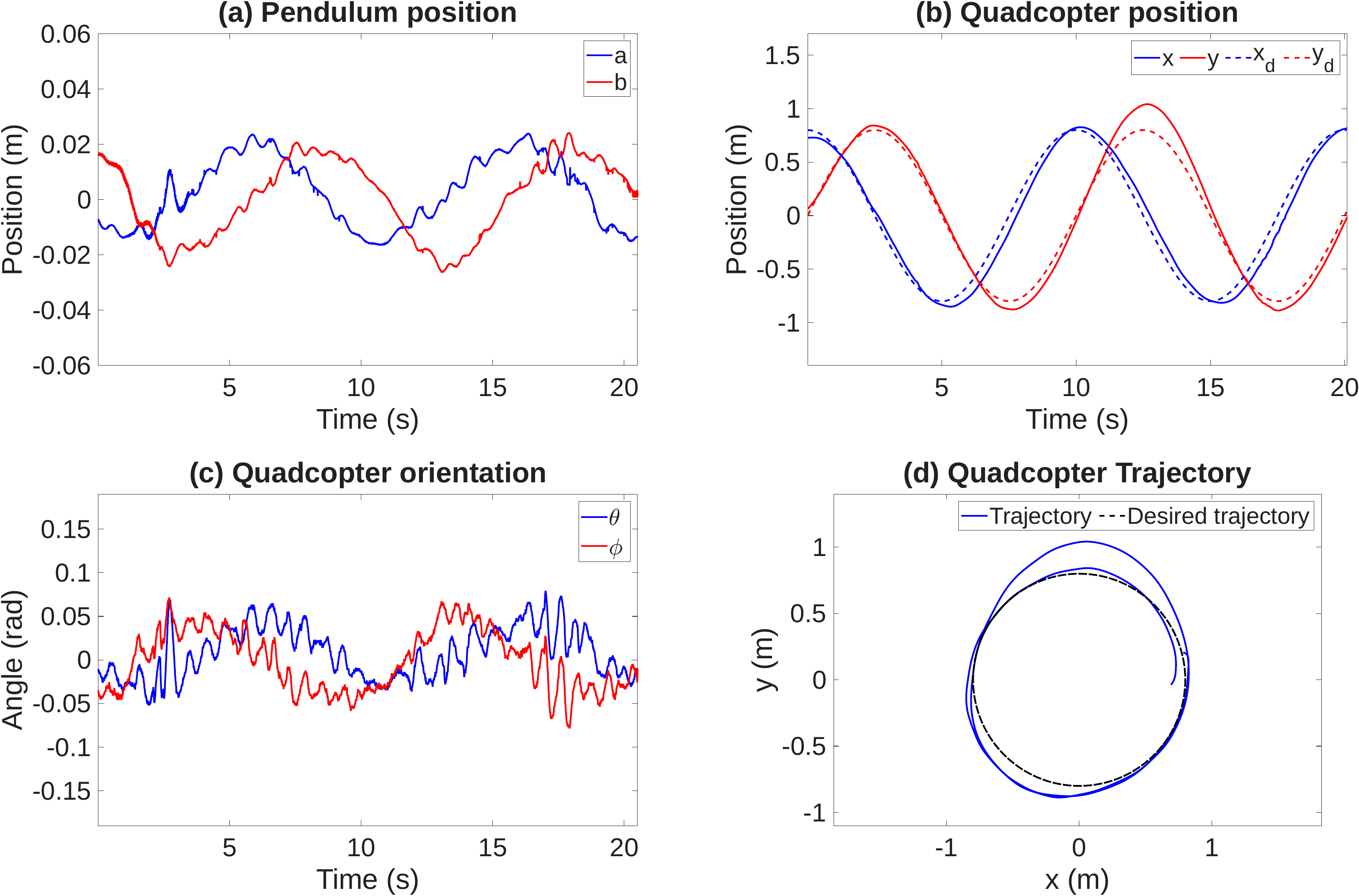}
    \caption{MPC controller performance while tracking a circular trajectory.}
    \label{fig:prac_circle}
\end{figure}

Fig.~\ref{fig:prac_circle} demonstrates sufficient tracking performance in the $x$ axis with minimal undershoot and a slight delay. However, the controller showed poorer performance in the $y$ axis with a maximum deviation of approximately $0.2$  m. The controller's ability to stabilize the pendulum was notable as the pendulum’s center of mass had a maximum deviation of $2.5$ cm from the quadcopter’s geometric center which translates to an angle of $2.87$ degrees from the vertical position. The large error in the $y$ axis shown in Fig.~\ref{fig:disturbance} and Fig.~\ref{fig:prac_circle} suggests that the cost coefficients for \(y\) and \(\dot{y}\) should be fine-tuned for a faster settling time and a smaller steady state error.

\subsection{Pendulum Length Estimation}
As the Genetic Algorithm requires a time-varying trajectory, the pendulum length was estimated while the quadcopter continuously tracked a circular trajectory of radius \(0.4\) m while balancing the pendulum.

\begin{figure}[h!]
    \centering
    \includegraphics[width=0.8\linewidth]{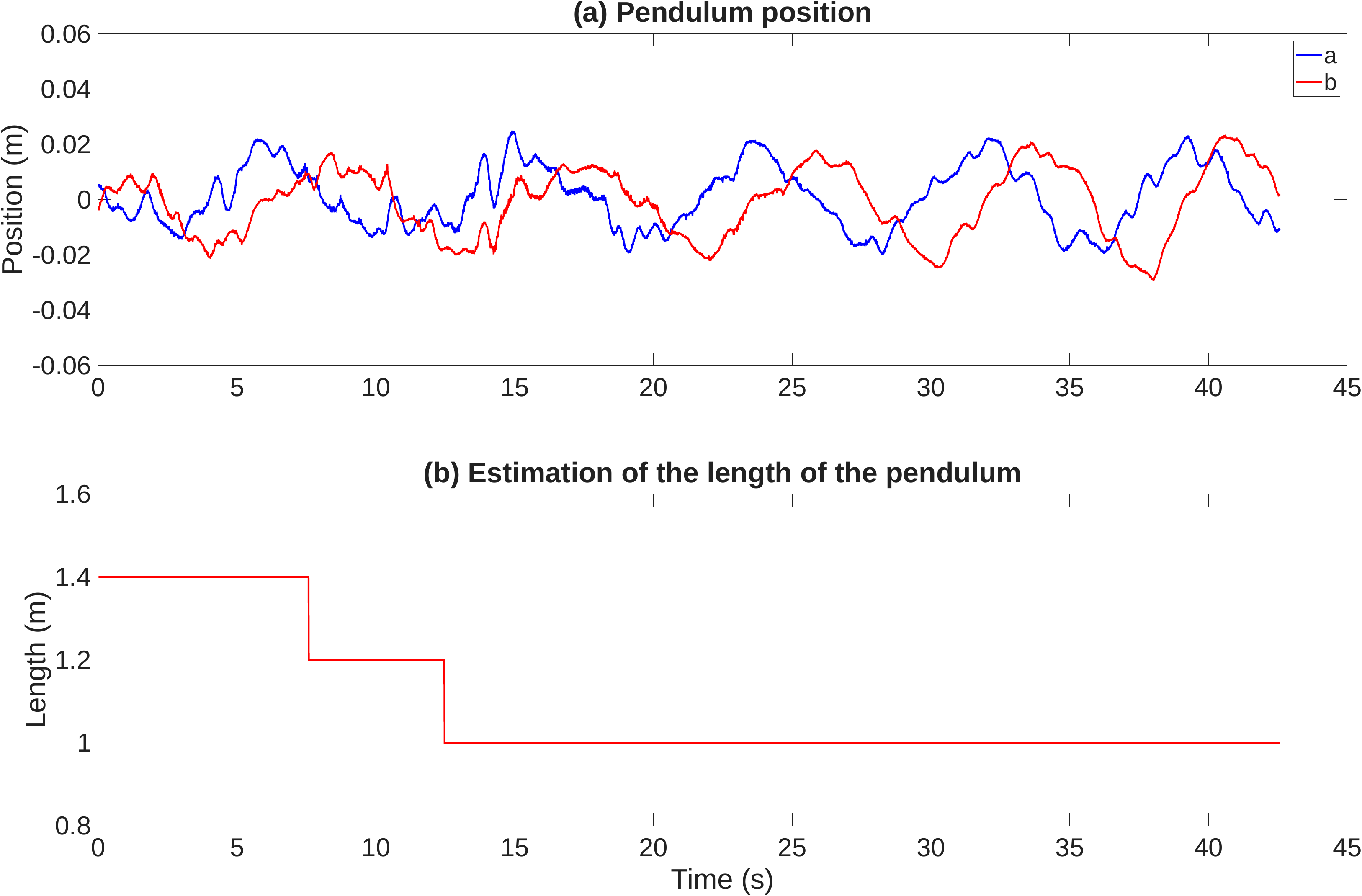}
    \caption{MPC controller performance when estimating pendulum length using a Genetic Algorithm Estimator when pendulum length is initially set to \(l=1.4m\) in the MPC prediction model.}
    \label{fig:ga_0_7}
\end{figure}
Fig.~\ref{fig:ga_0_7} shows that the Genetic Algorithm has proven to be effective when the pendulum length was initialized to $1.4$m in the MPC prediction model as the estimation of the pendulums length was able to converge to the pendulums true length within $12.43$ seconds. After sending the estimated pendulum length to the MPC algorithm the quadcopters trajectory tracking improved in the $x$ axis, however, trajectory tracking in the $y$ axis showed minimal improvement as the system response maintained a similar amplitude as the estimated value of $l$ continued to converge to the true value. 

\begin{figure}[h!]
    \centering
    \includegraphics[width=0.8\linewidth]{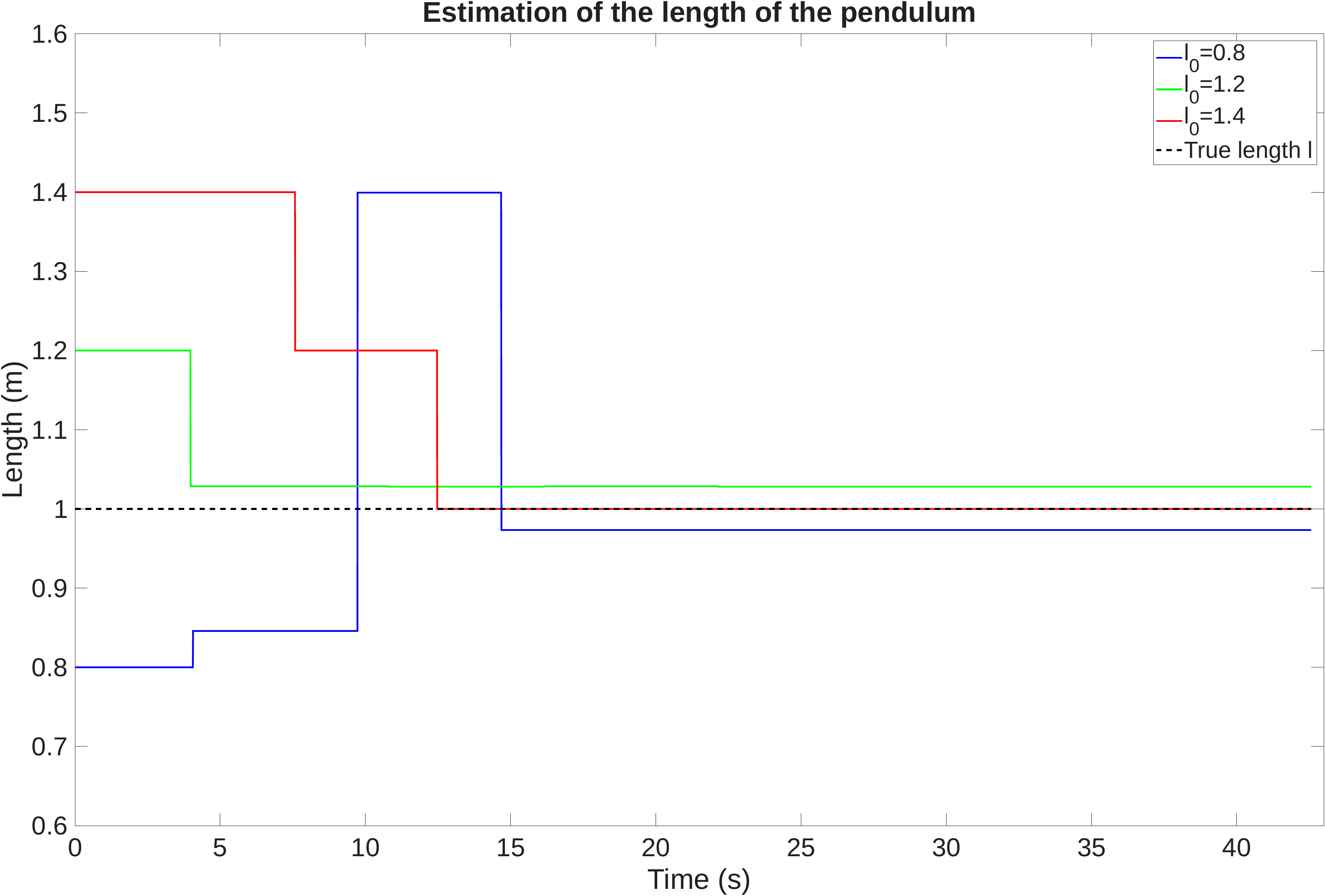}
    \caption{Genetic algorithm estimator convergence when the initial pendulum length is set to $0.8$, $1.2$ and $1.4$ metres in the proposed system.}
    \label{fig:allestimations}
\end{figure}

The Genetic Algorithm estimator was also tested and demonstrated to be effective when the pendulum length was initialized to $0.8$ and $1.2$ metres. Fig.~\ref{fig:allestimations} shows that the estimation of the pendulums length was able to converge to $0.973$ metres in $14.48$ seconds and $1.029$ meters in $3.99$ seconds respectively. All the pendulum length estimations and convergence times are summarized in Table \ref{tab: ga_results}.

\begin{table}[H]
\centering
\caption{Genetic Algorithm discretized estimation results and convergence speed when initial pendulum length was set to $0.8$, $1.2$, and $l=1.4$ metres.}
\label{tab: ga_results}
\begin{tabular}{l p{2cm} p{2.2cm}}
\toprule
\textbf{$l_0$} & \textbf{Convergence time (s)} & \textbf{Discrete error (m)} \\
\midrule
0.8                   & 14.68             & 0.027 \\
1.2                   & 3.99         & 0.029 \\
1.4                   & 12.43       & 0.0 \\
\bottomrule
\end{tabular}
\end{table}

Although the estimates of the pendulum length were highly accurate, the algorithm showed little to no convergence after the first $14$ seconds. This indicates that the algorithm did little exploration, which suggests the need for reducing the number of preserved individuals for the next generation and also for using a higher mutation to prevent premature convergence. It was observed that when the pendulum’s estimated length increases or decreases by a value greater than $0.8$m, the MPC algorithm’s control outputs fluctuate significantly, leading to instability. Furthermore, rapid fluctuations in length estimation occurred in the experimental process, this could be attributed to noisy data.



\section{Conclusion}
In this work, an MPC controller and an online GA estimator were proposed. The system was then implemented in the real world using a quadcopter and a motion capture system. Controller performance was demonstrated with disturbance rejection, circular-trajectory tracking, and pendulum length estimation trials. Future work could look to improve the GA response by gradually increasing or decreasing the estimated length in increments rather than instantly replacing the length with the new estimated length. Additionally, exploration of the search space could be refined by increasing the mutation rate and decreasing the number of preserved individuals for the best preservation method.

\bibliography{References} 
\end{document}